\documentclass[11pt,twocolumn,twoside]{IEEEtran}
\usepackage{amsmath}
\pdfoutput=1
\usepackage[margin=0.75in,headheight=0.45in]{geometry}
\usepackage[pdftex]{epsfig}
\usepackage{amsfonts}
\usepackage{amssymb}
\usepackage{fancyhdr}
\usepackage{graphicx}

\usepackage[utf8]{inputenc} 
\usepackage[T1]{fontenc}    
\usepackage{hyperref}       
\usepackage{url}            
\usepackage{booktabs}       
\usepackage{amsfonts}       
\usepackage{nicefrac}       
\usepackage{microtype}      
\usepackage{multirow}
\usepackage{amsmath}
\usepackage{subcaption}
\usepackage{xcolor}

\newcommand{\bfz}{\textbf{z}}
\newcommand{\bfx}{\textbf{x}}
\newcommand{\iu}{{i\mkern1mu}}

\newif\ifcomments
\commentstrue 
\ifcomments
    \providecommand{\jens}[2][]{{\protect\color{blue}{[Jens:\textbf{#1} #2]}}}
\else
    \providecommand{\jens}[2][]{}
\fi

\begin{document}
\title{\vspace{0.2in}\sc Generative Modeling of Atmospheric Convection}
\author{Griffin Mooers$^{1}$\thanks{Corresponding author: G Mooers, gmooers@uci.edu $^1$Department of Earth System Science, University of California at Irvine, CA, USA $^2$Department of Computer Sciences, University of California at Irvine, CA, USA
$^3$Department of Earth and Environmental Engineering, Columbia University, New York, NY, USA}, Jens Tuyls$^{2}$, Stephan Mandt$^{2}$, Mike Pritchard$^1$, Tom Beucler$^{1,3}$ \thanks{The authors thank Liran Peng for assistance with data generation as well as Karthik Mukkavilli and Ruihan Yang for helpful discussions that advanced this research project. GM is grateful to the MAPS Program and NSF grant 1633631 for funding support. TB is supported by NSF grants OAC-1835769, OAC-1835863, and AGS-1734164. SM acknowledges funding from DARPA (HR001119S0038). MP is also supported by NSF grant OAC-1835863. The code for this study can be found at \url{https://github.com/gmooers96/CBRAIN-CAM/tree/master/MAPS}}}

\maketitle
\thispagestyle{fancy}
\begin{abstract}

While cloud-resolving models can explicitly simulate the details of small-scale storm formation and morphology, these details are often ignored by climate models for lack of computational resources. Here, we explore the potential of generative modeling to cheaply recreate small-scale storms by designing and implementing a Variational Autoencoder (VAE) that performs structural replication, dimensionality reduction, and clustering of high-resolution vertical velocity fields. Trained on $\sim6\cdot \mathrm{10^{6}}$ samples spanning the globe, the VAE successfully reconstructs the spatial structure of convection, performs unsupervised clustering of convective organization regimes, and identifies anomalous storm activity, confirming the potential of generative modeling to power stochastic parameterizations of convection in climate models.
\end{abstract}

\section{Motivation}
Boxed in by computational limits, many of the details of our atmosphere remain too minute to explicitly resolve in climate models~\cite{doi:10.1175/BAMS-84-11-1547, doi:10.1029/2019MS001610, schneider_17}. Key physics driving convection and cloud formation occur on the scale of meters to a few kilometers, while typical modern climate models have a resolution of $100-200 \mathrm{km}^2$ horizontally - meaning important sub-grid processes are parameterized. Computational capabilities are advancing, and climate models are increasingly common, in particular those with three-dimensional explicit resolution of clouds systems. However, the capability to run these models for the $\sim $100-year timescales needed is often impractical~\cite{Jensen2041, doi:10.1175/JCLI-D-12-00695.1, Bjorn_2015} and the information content they generate about the details of cloud and storm organization are frequently overwhelming to analyze at its native scale. This has left significant gaps in knowledge about many of the details of cloud-climate feedbacks and the relationship between storm organization and its thermodynamic environment~\cite{doi:10.1175/BAMS-84-11-1547, Bjorn_2015}. However, deep learning, and in particular generative models, may provide a path to a better understanding of these phenomena and their role driving the weather and climate of our world.
\par

The application of machine learning in the physical sciences has increased exponentially in recent years but with important avenues still largely unexplored. In climate modeling, deep neural networks have been re-purposed to emulate the large-scale consequences of storm-level heating and moistening over the atmospheric column to replicate mean climate and expected precipitation patterns and extremes~\cite{Rasp9684, doi:10.1029/2018GL078202, Bjorn_2015, doi:10.1029/2018MS001390, doi:10.1029/2018MS001351}. However, much of this work has been confined to deterministic neural networks that ignore the interesting stochastic details of eddy and storm organization. The recent application of Generative Adversarial Networks (GANs,~\cite{goodfellow2014generative}) to the Lorenz '96 Model suggests a potential, under-explored role for generative models in atmospheric sciences -- particularly towards stochastic parameterizations~\cite{doi:10.1029/2019MS001896, Loenze_96_GAN_2}. There have also been initial successes using various types of GAN architectures to generate plausible Rayleigh-Bernard convection. In particular, adding informed physical constraints to GAN loss functions seem to improve the generation of these non-linear fluid flow systems~\cite{Yang2020PhysicsInformedGA, WU2020109209, Stat_Stinis, Yang_Physics}. While promising, such techniques have thus far been restricted to idealized turbulent flows of reduced dimension and complexity; there is ample room to explore generative modeling methods for representing convective details amidst settings of realistic geographic complexity. Meanwhile, generative modeling besides GANs have not been as thoroughly considered for turbulent flow emulation and could potentially power climate models down the line. 
\par
VAEs may prove more appropriate than GANs for these climate applications given their design containing both a generative and representational model, their often superior log-likelihoods and reconstruction simulations, and practical advantages including stabler training results, easier performance benchmarking, and more interpretable latent manifold representations~\cite{Decoder_Based_Gen, Advar_Variational_Bayes, IntroVAE}. Modified VAEs can reconstruct plausible two-dimensional laminar flow with computational efficiency beyond what is common when numerically solving linear differential equations~\cite{navier_VAE}. There has been preliminary work using VAEs for the clustering of atmospheric dynamics -- a gain again relying on simplified Lorenz '96 model data as well as potential vorticity fields and geopotential heights~\cite{SupernoVAE, Polar_Vortex}. This application of representation learning across a variety of simplified simulations suggests VAEs offer great potential as both an engineering tool to help escape computational limits on the generative side and may provide the ability to learn and extract hidden organizational details in atmospheric dynamics on the representation side. However, to the best of our knowledge, this is the first study to use a VAE for representational learning on the details of convective organization and associated gravity wave radiation\footnote{Here we are referring to internal gravity waves, which are horizontally-propagating disturbances in the atmosphere generated by density perturbations, e.g. from deep convection, frontogenesis, or topography.} as revealed by spatial snapshots of vertical velocity -- an inherently chaotic and bimodal variable~\cite{doi:10.1029/2009JD013091} -- across a dataset large enough to nonetheless encompass the spatiotemporal diversity of turbulence regimes in the atmosphere. As far as we know, this is also the first study to constrain a VAE's output statistics by adding a covariance constraint term to its loss function to improve representation and capture variance details at small spatial scales in the turbulent atmospheric boundary layer, which can be considered one of the most difficult locations for climate models. Our results demonstrate the power of VAEs to accurately reconstruct high-resolution climate data, as well as their ability to leverage dimensionality reduction for high level feature learning and anomaly detection. This casts VAEs as promising tools for both dynamical analysis and stochastic parameterization of fine-scale atmospheric processes from cloud-resolving data. 

\section{Method}

In this Section, we discuss the architecture of the three machine-learning models used here, the design of our covariance constrained VAE loss function, and the generation and preprocessing of the atmospheric simulation data.

\subsection{Architecture}

\begin{table}[tb]
\begin{tabular}{lcccc}
 \toprule 
 \textbf{Layer} & \textbf{Filters} & \textbf{Kernel} & \textbf{Stride} & \textbf{Activation} \\
 \midrule 
 2D Conv & 64    &3x3&   2 & relu \\
 2D Conv  & 128    &3x3&   2 & relu \\
 2D Conv  & 512    &3x3&   2 & relu \\
 2D Conv ($\mu$)  & 64    &3x3&   2 & relu \\
 2D Conv ($\sigma$) & 64    &3x3&   2 & relu \\
 \bottomrule
\end{tabular}
\caption{Our Encoder architecture. Conv refers to a convolutional hidden layer. The first hidden Conv layer receives an input vector of 32x128 (30x128 expanded by padding) representing a vertical velocity snapshot.}
\label{tab:encoder}
\end{table}

\begin{table}[tb]
\begin{tabular}{lcccc}
 \toprule 
 \textbf{Layer} & \textbf{Filters} & \textbf{Kernel} & \textbf{Stride} & \textbf{Activation} \\
 \midrule
 2D Conv-T  & 1024    &3x3&   2 & relu\\
 2D Conv-T  & 256    &3x3&   2 & relu\\
 2D Conv-T  & 64    &3x3&   2 & relu\\
 2D Conv ($\mu$)  & 1    &3x3&   2 & sigmoid\\
 2D Conv ($\sigma$) & 1    &3x3&   2 & linear\\
 \bottomrule
\end{tabular}
\caption{Our Decoder architecture. Conv-T refers to a transposed convolutional hidden layer.}
\label{tab:decoder}
\end{table}

Our VAE takes vertical velocity fields formatted as (30$\times $128) 2D images. We adopt a fully convolutional design\footnote{Earlier experiments used architecture similar to models used for CIFAR-10 data~\cite{cifar} with fully connected dense layers separating the encoder and the decoder from the latent space, but led to discouraging reconstructions plagued by posterior collapse and an inability to represent the spatial patterns of convection.} to preserve local information, which is essential in atmospheric convection modeling (Tables~\ref{tab:encoder} and~\ref{tab:decoder}). We obtain meaningful reconstruction performance by ensuring that the information bottleneck in the VAE is not too severe, i.e. that the latent space is still wide enough to preserve enough fine features of the vertical velocity fields (in our case of dimension 1024), and by implementing annealing techniques outlined in~\cite{Higgins2017betaVAELB, Alemi2018FixingAB}. Here, we analyze two successful VAEs: One with a traditional negative ELBO in the loss, and one with an additional covariance constraint in the loss. As a baseline, we also implemented a regular autoencoder of the same design as above, with two key differences: All activations were replaced with the identity function and our covariance constrained loss was replaced with the mean-squared error. We refer to this model as the ``linear'' model, and use it to better quantify the added value of VAEs for modeling atmospheric convection.  

\subsection{VAE Loss Implementation\label{sub:VAE_Loss}}

The total loss is the sum of two terms: the negative of the Evidence Lower Bound (ELBO), commonly used as the total VAE loss, and a covariance constraint loss term~\cite{navier_VAE, Stat_Stinis, autoencoder} on the covariance matrix that we weigh by $ \lambda\in {\mathbb R^{+}}$:

\begin{equation}
    \mathrm{Loss}\overset{\mathrm{def}}{=}-\mathrm{ELBO}+\lambda\times{\mathrm{CC}},
\end{equation}


where CC is a ``covariance constraining'' term using the Frobenius norm $||\cdot||$ to measure the distance between the covariance, $\Sigma $, of the likelihood $p_\theta(\bfx | \bfz)$ and the covariance, $\Sigma $, of the true data distribution $p(\bfx) $.  $\mathrm{\theta}$ refers to model parameters and $\mathrm{\bfx}$ refers to observed vertical velocity fields:

\begin{equation}
    \mathrm{CC}=||\Sigma(p_{\theta}(\bfx|\bfz)) - \Sigma(p(\bfx))||.
    \label{eq:CC}
\end{equation}

Unconstrained VAEs ($\lambda=0$), henceforth referred to as ``VAE'' for short, maximize the ELBO, defined as the sum of the log-likelihood  $p_{\theta}(\bfx|\bfz)$, and the Kullback-Leibler (KL) Divergence between $p(\bfz) $ and  $q_\phi(\bfz | \bfx)$:
\begin{equation}
    \label{eq:elbo}
    \begin{aligned}
    \mathrm{ELBO}(\bfx; \theta, \phi, \bfz) & = \mathbb{E}_{q_\phi(\bfz | \bfx)}[\log p_\theta(\bfx|\bfz)] \\
     & - D_{\text{KL}}(q_\phi(\bfz | \bfx) \,||\, p(\bfz)), 
    \end{aligned}
\end{equation}

where $\mathrm{\phi}$ are our variational parameters which are learned jointly with the model parameters, $\mathrm{\theta}$. $p(\bfz)$ refers to the prior and $q_\phi(\bfz | \bfx)$ refers to the estimated posterior. We denote hidden variables as $\mathrm{\bfz}$. Minimizing the KL loss term regularizes the variational parameters in the model and makes the VAE posterior more similar to the VAE prior. Maximizing the log-likelihood enables the VAE to produce realistic vertical velocity fields where the output will be more closely aligned with the latent variable of the model. Following~\cite{Kingma2014AutoEncodingVB}, we assume that the prior over the parameters and the hidden variables are both centered isotropic Gaussian and calculate ELBO using equation (24) of~\cite{Kingma2014AutoEncodingVB}.

To control the rate-distortion trade-off~\cite{Alemi2018FixingAB}, we implement linear annealing to the KL loss term following~\cite{Bowman2016GeneratingSF}, where the KL term is multiplied by an annealing factor linearly scaled from 0 to 1 over the course of training. In our VAE, linear annealing results in significantly lower KL losses and more interpretable latent spaces. 

Finally, to generate vertical velocity fields with realistic spatial variability, we additionally implement covariance-constrained VAEs. Following Equation \ref{eq:CC}, the covariance constraint is defined as the Frobenius norm of the covariance matrix error, which we estimate over each batch during optimization. We choose a pre-factor $\mathrm{\lambda=10^{6}}$ so that the magnitude of the covariance constraint matches that of the reconstruction loss, resulting in a covariance-constrained VAE ``CC-VAE'' that generates more faithful covariance matrices.

\subsection{Data \& Preprocessing}\label{Dataset}

\subsubsection{Cloud-Resolving Data}

To train and test our VAE, we rely on snapshots of vertical motions with explicitly-resolved moist convection and gravity wave radiation obtained from $\sim$15k instances of a Cloud-Resolving Model (CRM)~\cite{10.1175/1520-0469(2003)060<0607:CRMOTA>2.0.CO;2, Khairoutdinov1999ALE} embedded within a host Global Climate Model (GCM). The CRMs operates at a 20s native timestep data and we extract state snapshots from it every 15 minutes, the frequency with which its horizontal average state is permitted to interact with its host GCM. We perform a 100-day multi-scale climate simulation to generate data showing details of atmospheric convection within a tropical belt from 20N to 20S latitudes. Specifically, at each $1.9^{\circ} \times 2.5^{\circ}$ horizontal grid cell of the Super-Parameterized Community Atmosphere Model (SPCAM5), we embed a 128-column System for Atmospheric Modeling (SAM) micro model with kilometer scale horizontal resolution; both the host and embedded models use 30 vertical levels. This entire dataset comes to a size of 1.3 Tb. For our purposes, there is 30 level by 128 CRM-column "snapshot" or "image" of a convective-scale vertical velocity field at each latitude-longitude grid cell that we feed into the encoder of our neural network. We train our VAEs on sub-samples of this data staged on UC Irvine's GreenPlanet Super-computing node and our machine learning simulations are powered by two NVIDIA Tesla V100 and one NVIDIA Tesla T4 GPUs.

\subsubsection{Preprocessing}

To reduce data volume for efficient training and to ensure our VAE is exposed to a plethora of convective motion, we selectively sample from the initial 1.3Tb SAM dataset. We restrict our initial data volume to the 144 latitude/longitude coordinates with a detectable diurnal cycle of precipitation where amplitude of daily precipitation is greater than two times its standard deviation within the larger-scale host model. This precipitation filtering ensures samples of strong convection get placed into the training dataset, as a persistent diurnal cycle of precipitation often indicates deep convection and the presence of mesoscale convective systems~\cite{10.1175/MWR3467.1}. Within these selected grid cells, the vertical velocity values range from $37.3 \mathrm{m\ s^{-1}}$ to $-17.4 \mathrm{m\ s^{-1}}$ and are then scaled from 0-1 by subtracting the minimum and dividing by the range. 

We shuffle data in the spatial and temporal dimensions prior to training. We use An 80$\%$/20$\%$ training/test split for all models. To ensure a balanced dataset of different convective types, we apply K-means clustering with two centroids to group data with active and inactive vertical velocity fields. We then sample equally from both clusters without replacement to design a balanced dataset for the VAE. This new 4.3Gb dataset has a 111206/27802 training/test split. Since the horizontal domain is doubly-periodic, two vertical velocity updrafts of equal magnitudes and size located at different horizontal locations are physically identical. To prevent the VAE from treating them as different at the expense of reconstruction magnitude and variance, we preprocess all samples so that the center of the vertical velocity field is the location of strongest convection present in the sample. We define the ``strongest convection'' as the largest absolute value of spatially-averaged vertical velocity, from 400hPa to 600hPa in the vertical and using a moving average of 10km horizontally.  

\subsection{Quantifying Reconstruction Performance}

We quantify the reconstructions of our final VAE and CC VAE as well as our linear baseline using the following metrics:

\subsubsection{Hellinger Distance} We calculate the Hellinger distance H between the discrete distributions to gauge similarity~\cite{10.5555/2380985}:

\begin{equation}\label{eq:hellinger}
H(p,q) = \sqrt{\sum_{i=1}^k \frac{(\sqrt{p_i}-\sqrt{q_i})^2}{2}} 
\end{equation}

where $\mathrm{p}$ is the distribution of the original vertical velocity fields and $\mathrm{q}$ is the distribution of the corresponding reconstruction.

\subsubsection{Mean Squared Error (MSE)} To provide an overall skill of the reconstruction, the MSE is calculated between each original sample and its corresponding reconstruction.

\subsubsection{Spectral Analysis} To better understand the skill of the VAE reconstruction from a spatial perspective, we perform one-dimensional spectral analysis on each sample and reconstruction at all 30 levels in the vertical dimension. We examine four vertical levels commonly used in meteorology: 850hPa (top of the boundary layer), 700hPa (lower troposphere), 500hPa (mid-troposphere), and 250hPa (upper-troposphere) to see how our VAEs capture the spatially-resolved vertical velocity variance throughout the atmosphere. We calculate the power spectral density $\mathrm{\Phi_k}$ using: 

\begin{equation}
\Phi_k \overset{\mathrm{def}}{=}\frac{\Delta n}{N}\left|\sum_{j=0}^{N-1} y_je^{\frac{-\iu jk}{NT}}\right|^2 
\end{equation}

where $\mathrm{N}$ is the length of the x dimension, $\mathrm{y_j}$ is the sample or reconstruction, $\mathrm{T}$ is 1/length, $\iu$ is the imaginary unit and $\mathrm{k}$ is the vertical level of interest in hPa (850, 700, 500, or 250)~\cite{Cooley1965AnAF}.

\section{Result \& Discussion}

\begin{table}[tb]
\begin{tabular}{lccc}
 \toprule 
 \textbf{Model} & \textbf{MSE} & \textbf{Hellinger Distance} & \textbf{Frobenius Norm}  \\
 \midrule
 Linear  & 4.2e-6    &2.0e-3&   8.0e-3\\
 VAE  & 1.1e-5    &3.1e-4&   3.2e-4\\
 CC VAE  & 4.5e-6    &2.0e-3&   8.0e-6\\
 \bottomrule
\end{tabular}
\caption{\textbf{Quantitative Reconstruction Metrics.} We compute the MSE and Hellinger Distance between true and predicted reconstructions. This shows the baseline is equally good at predicting the mean reconstruction. We also compute the Frobenius Norm of the error in the covariance matrices of the true data and the reconstructions. Both VAEs capture more of the covariance structure of the data than the linear baseline.}
\label{tab:mse_hellinger}
\end{table}

\begin{figure*}
    \begin{centering}
        \includegraphics[width=\textwidth]{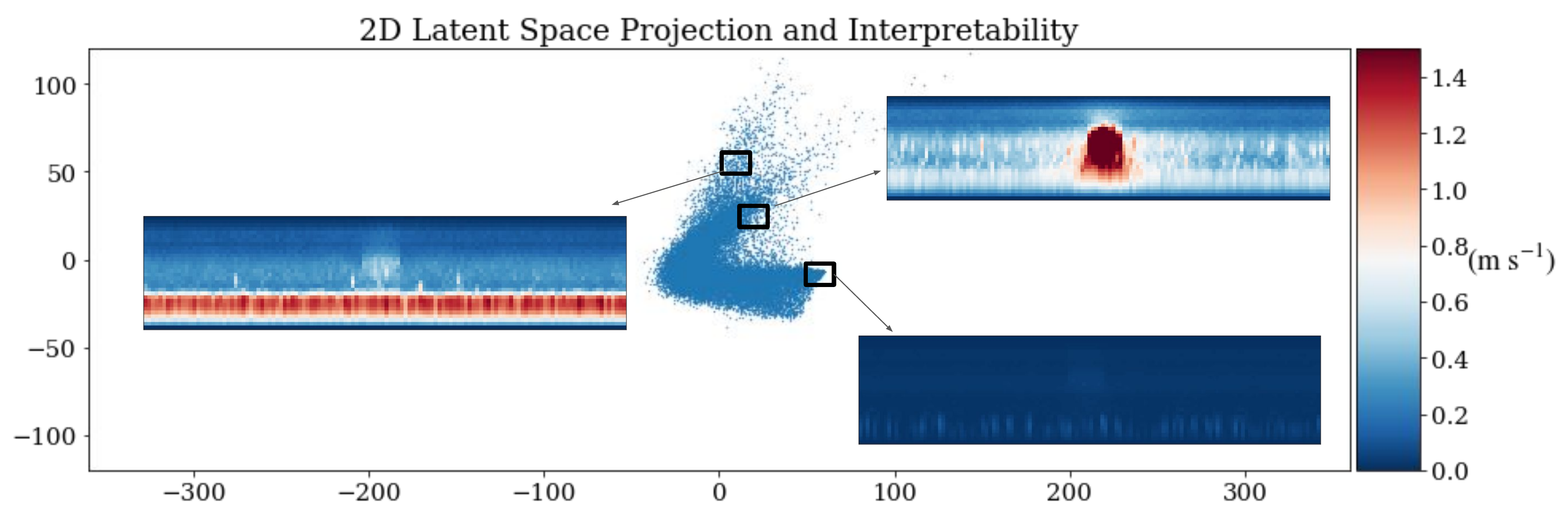}
    \par\end{centering}
  \caption{\textbf{Visualization of the latent space} originally in dimension 1024, but reduced to dimension 2 by Principle Component Analysis (PCA)~\cite{pearson_karl_1901_1430636}. The standard deviations of different types of convection the VAE learns to cluster are embedded near corresponding clusters. This suggests the VAE learns an interpretable clustering of the data, with means and variances both contributing to the results.}
  \label{fig:latent_space_analysis}
\end{figure*}

\begin{figure*}
  \begin{centering}
    \includegraphics[width=\textwidth]{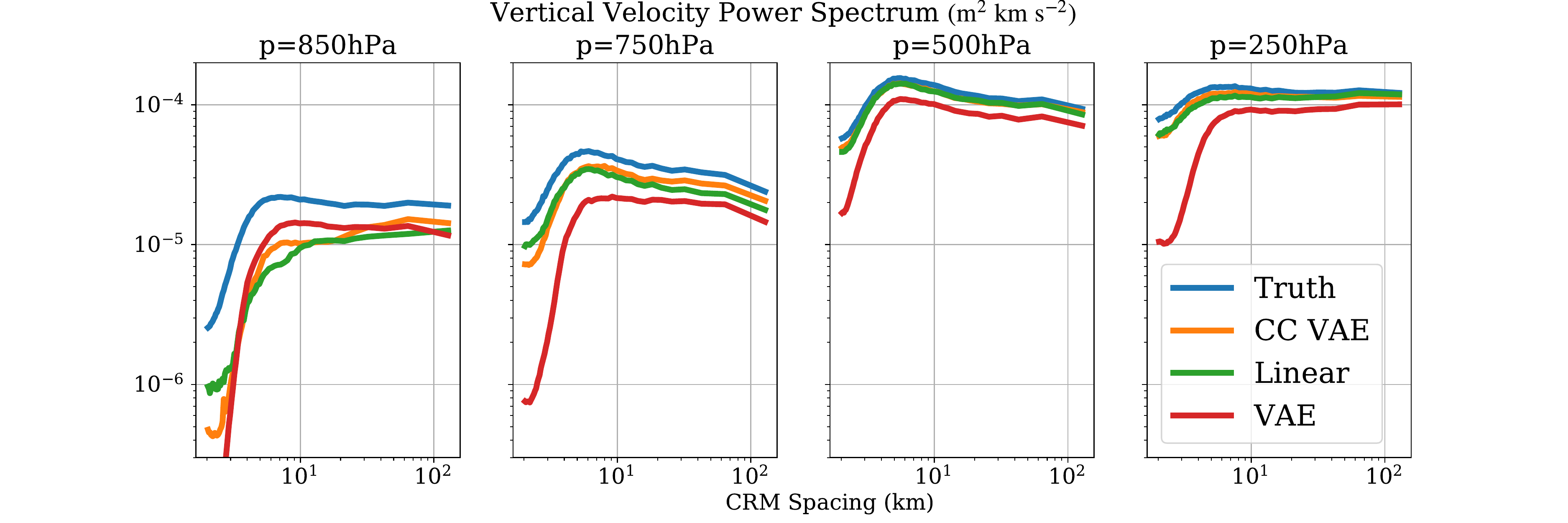}
  \end{centering}
  \caption{\textbf{Spectral Analysis} at 4 different levels of the atmosphere comparing the test data to our best VAE and CC VAE as well as a linear model. At small spatial scales we see the importance of the Covariance Constraint to capture the variance native to convection (orange vs. red).}
  \label{fig:spectral_analysis}
\end{figure*}

\begin{figure*}
  \begin{centering}
    \includegraphics[width=0.75\textwidth]{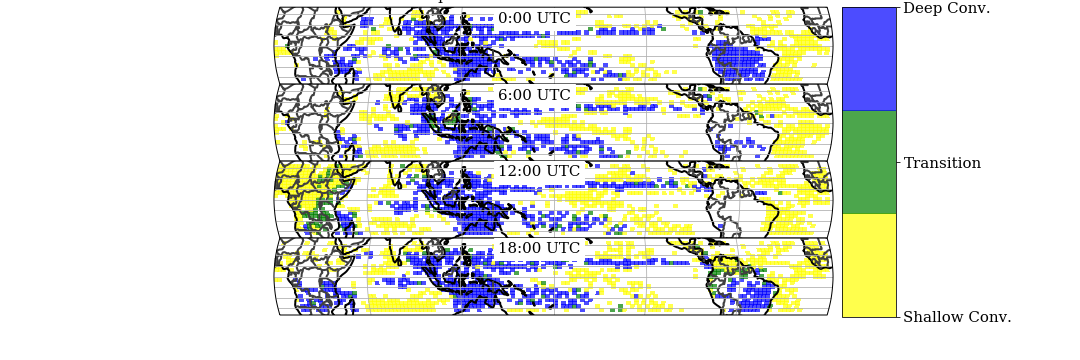}
  \par\end{centering}
  \caption{\textbf{Convection Type Predictions} The diurnal composite from a ten day average at four unique times of day are shown above. The VAE predicts the type of convection occurring in tropical locations over the course of a typical Boreal Winter Diurnal Cycle. Blue coloring refers to a VAE prediction of deep convection, yellow to a VAE prediction of shallow convection and green to a convective type transitioning between shallow or deep convection. Areas where the VAE detects little convection are blanked out. Semantic similarities of the VAE latent space are reflected in the global geospatial weather patterns.}
  \label{fig:diurnal_composite}
\end{figure*}

\begin{figure*}
  \begin{centering}
    \includegraphics[width=0.85\textwidth]{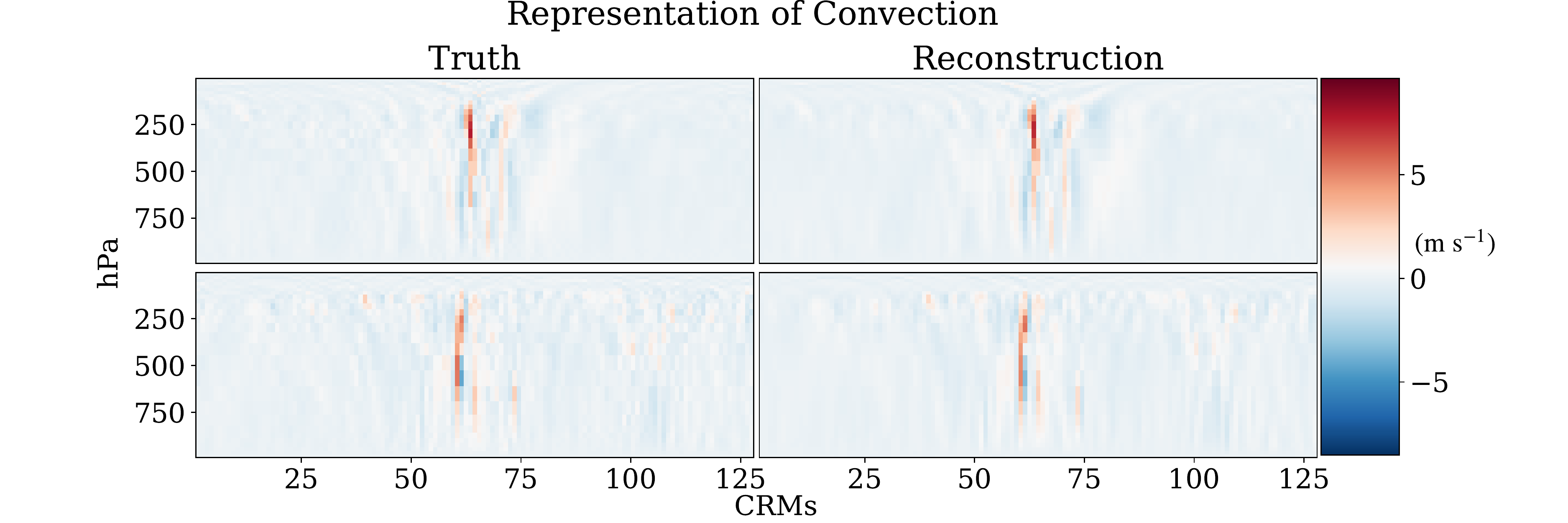}
  \par\end{centering}
  \caption{\textbf{Reconstructions} The trained VAE reconstructions closely resemble those from the test dataset and accurately predict the location, magnitude and spatial structure of convective plumes.}
  \label{fig:Reconstructions}
\end{figure*}

Our VAE trained on cloud-resolving climate data produces accurate vertical velocity field reconstructions. When we provide the high resolution training dataset and appropriate convolutional architecture, our VAE learns remarkably accurate representations of any type of convection found within the test dataset. Our VAE captures the magnitude, proper height, and structure across deep convective regimes, shallow convective regimes, and non-convecting regimes  (Figure~\ref{fig:Reconstructions}). When the ``Covariance Constraining'' term is added to create a physically informed loss, the CC VAE performance improves enough to match a linear baseline (Table~\ref{tab:mse_hellinger}). But unlike many other image recognition tasks generative models perform, reconstructing the mean of the convection is necessary but not sufficient -- we must capture the variance and correlation in the vertical velocity fields. The CC VAE reconstructs variance better than a traditional convolutional VAE and at least on par with the linear baseline  (Table~\ref{tab:mse_hellinger}, Figure~\ref{fig:spectral_analysis}). Our CC VAE is the most versatile of our models with an accurate reconstruction performance overall at different levels of the atmospheric column and different convective spatial scales based on the power spectra of the three models (Figure~\ref{fig:spectral_analysis}). This precision across both small and large spatial scales revealing our CC VAEs ability to emulate both the overall large pattern of convective plumes and the details within convective composition.
Our CC VAEs results replicate disparate structures of convection in both areas of high stochasticity near the atmospheric boundary layer, characteristic of shallow convection, as well as in the upper troposphere, where deep convective regimes dominate. At this stage CC VAEs match the performance of our linear baseline but do not exceed it.

\begin{figure*}
  \begin{centering}
    \includegraphics[width=0.85\textwidth]{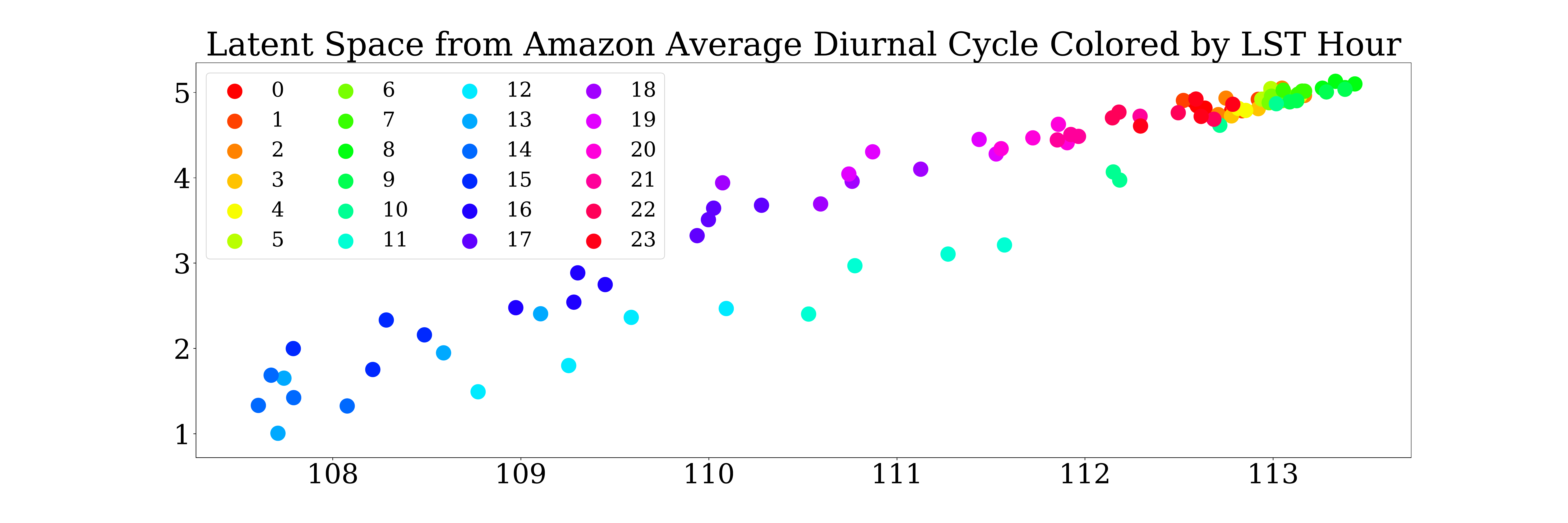}
  \par\end{centering}
  \caption{\textbf{2D PCA Temporal Projection} All spatial locations comprising the Amazon Rainforest are averaged together from November to February to get a single composite diurnal cycle that is fed through our trained VAE. The colors above correlate to time of day (Local Solar Time). The results show a clear separation in representation on the latent space of the timing of deepest convection and maximum precipitation (mid afternoon) from when shallow convection and calmer conditions dominate (early morning).}
  \label{fig:Amazon}
\end{figure*}

However, unlike the linear baseline, our VAE and CC VAE discover the details of convective organization by representation learning via dimensionality reduction and feature extraction. A 2D, deterministic PCA projection of our CC VAE latent space clusters and separates different convective types (Figure~\ref{fig:latent_space_analysis}). In particular, the distinction between deep and shallow convective regimes and non-convective regimes is encouraging (Figure~\ref{fig:latent_space_analysis}, please visit \href{https://tinyurl.com/yydhsrk6}{this link} for a complete animation of the 2D Projection of the latent space). The physical knowledge represented in our CC VAEs latent space stands alone from other forms of dimenionality reduction (PCA and t-SNE on the preprocessed data) where there is no evidence of distinction based on convective type. Furthermore, CC VAE predictions of convective type based solely on latent space location map back to a physically sensible pattern over the tropics with deep convection concentrated on land over the Amazon and African Rainforests as well as over the Pacific Warmpool (Figure~\ref{fig:diurnal_composite}). These predictions from latent space location not only map convection type in a spatially coherent pattern, but also capture the change in convection type with the diurnal cycle over moist, tropical continents (Figure~\ref{fig:diurnal_composite}, please visit \href{https://tinyurl.com/y43w2rmm}{this link} for a complete animation of the tropical diurnal cycle). When we exclusively restrict the test dataset to an Amazon Diurnal Composite, the known coherent transitions from shallow to deep convection that occur over tropical rain-forest in response to solar heating of the diurnal cycle correspond to monotonic trajectories in the latent space projection, verified using both t-SNE and PCA (Figure~\ref{fig:Amazon}). Further tests are required on more complex convective transitions to understand the extent of the physical meaning of the CC VAE latent space, but these initial positive results suggest great potential for physically constrained VAEs as a tool in atmospheric dynamics to uncover information about convective transitions, storm morphology and propagation. 

We also evaluate ELBO (Equation~\ref{eq:elbo}) for each sample of our test data to find unusual storm development and activity in the dense CRM data. 

\begin{figure*}
  \begin{centering}
    \includegraphics[width=0.85\textwidth]{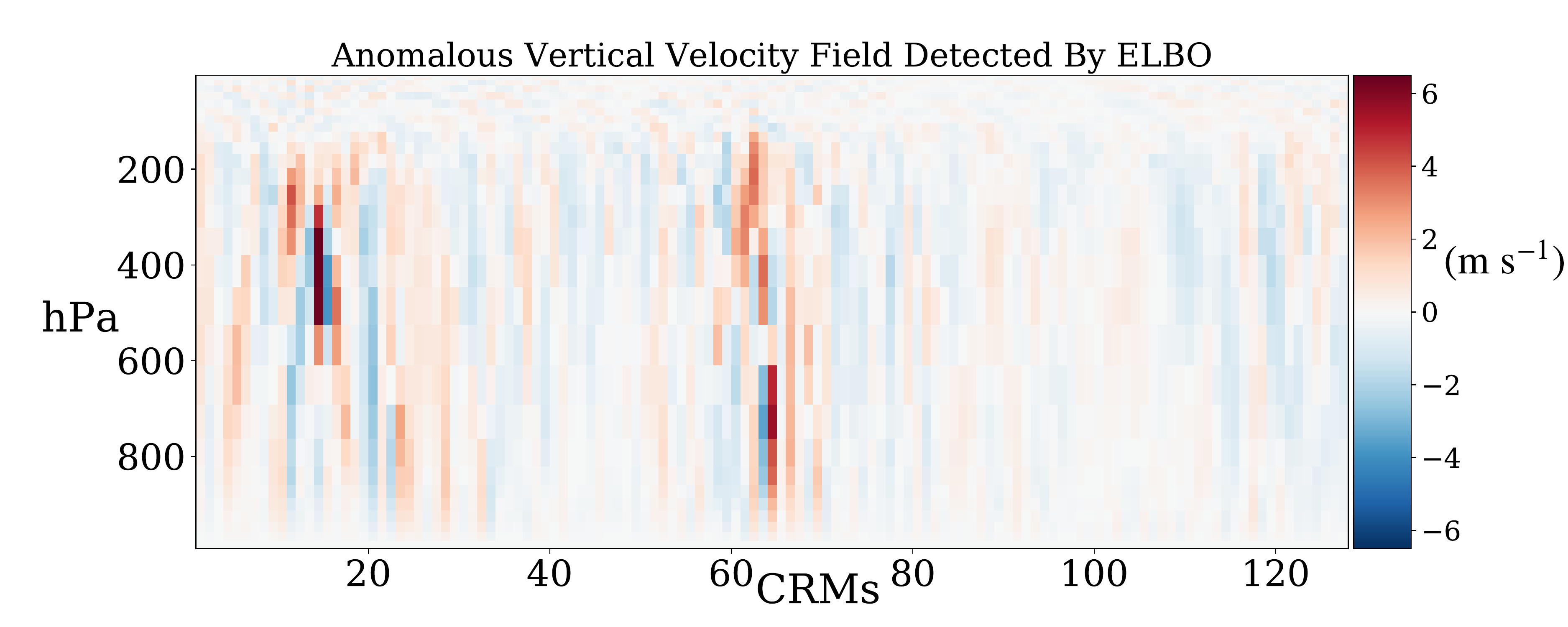}
  \par\end{centering}
  \caption{\textbf{Anomaly Detection} We use the ELBO in the VAE Loss function to identify the most anomalous vertical velocity fields. We show the 9th most anomalous field because it exhibits multiple deep convective plumes.}
  \label{fig:ELBO}
\end{figure*}

ELBO allows us to determine the degree to which a vertical velocity field, drawn from our models latent variables is an aberration in the data. Our VAEs inherent ability to detect anomalies in the vertical velocity data proves to be an elegant way to identify deep convection in a more thorough manner than traditional vertical velocity thresholding. An example of one such anomaly we identify is Figure~\ref{fig:ELBO} -- in this case an instance of two moderate storms developing in one CRM array. This phenomena would be less straightforward to locate through conventional methods, particularly given the size and density of data involved. Our VAEs attribute of anomaly detection learns characteristics of the data instead of naively thresholding based on priors experiences that may not reflect the composition of the dataset. This feature provides the potential to help identify interesting and unexpected weather phenomena from noise -- artifacts that might otherwise never be studied in overwhelmingly large and rich datasets.   

\section{Conclusion}

We develop a VAE to reconstruct immaculate convection images from a high-resolution, cloud-resolving dataset. Our VAE, particularly once a statistically constrained loss function is added, captures the variance and magnitude of distinct convective regimes. The latent space of the VAE proves to be a potent tool for making physically sensible predictions of convection type that accurately reflect the tropical atmosphere and capture the effects of solar heating through the diurnal cycle. The unique VAE loss function allows us to use ELBO to find anomalous storm development in a dense, high resolution dataset that traditional methods might miss. But there is much work to be done before a VAE could be implemented to power stochastic parameterizations for a climate model, likely requiring to condition the VAE on large-scale thermodynamics via expansion of the input vector. If successful, the ability to quickly and efficiently generate synthetic, detailed vertical velocity fields to help run climate models would be a valuable resource for the atmospheric sciences and meteorology communities. But improvements in the generative capabilities would likely come at the expense of the representation learning and the VAEs diagnosis of the physics of convection. We believe these preliminary physical intuitions achieved via representation learning represent a promising avenue for the broader application of generative modeling for advancing the field of atmospheric dynamics~\cite{Alemi2018FixingAB, Higgins2017betaVAELB} and warrant further investigation to understand their full potential.



\bibliographystyle{ieeetr}
\bibliography{main}

\begin{thebibliography}{10}

\bibitem{doi:10.1175/BAMS-84-11-1547}
D.~Randall, M.~Khairoutdinov, A.~Arakawa, and W.~Grabowski, ``Breaking the
  cloud parameterization deadlock,'' {\em Bulletin of the American
  Meteorological Society}, vol.~84, no.~11, pp.~1547--1564, 2003.

\bibitem{doi:10.1029/2019MS001610}
T.~R. Jones, D.~A. Randall, and M.~D. Branson, ``Multiple-instance
  superparameterization: 1. concept, and predictability of precipitation,''
  {\em Journal of Advances in Modeling Earth Systems}, vol.~11, no.~11,
  pp.~3497--3520, 2019.

\bibitem{schneider_17}
T.~Schneider, J.~Teixeira, C.~Bretherton, F.~Brient, K.~Pressel, C.~Schär, and
  A.~Siebesma, ``Climate goals and computing the future of clouds,'' {\em
  Nature Climate Change}, vol.~7, pp.~3--5, 01 2017.

\bibitem{Jensen2041}
E.~J. Jensen, G.~Diskin, R.~P. Lawson, S.~Lance, T.~P. Bui, D.~Hlavka,
  M.~McGill, L.~Pfister, O.~B. Toon, and R.~Gao, ``Ice nucleation and
  dehydration in the tropical tropopause layer,'' {\em Proceedings of the
  National Academy of Sciences}, vol.~110, no.~6, pp.~2041--2046, 2013.

\bibitem{doi:10.1175/JCLI-D-12-00695.1}
H.~Kalesse and P.~Kollias, ``Climatology of high cloud dynamics using profiling
  arm doppler radar observations,'' {\em Journal of Climate}, vol.~26, no.~17,
  pp.~6340--6359, 2013.

\bibitem{Bjorn_2015}
B.~Medeiros, B.~Stevens, and S.~Bony, ``Using aquaplanets to understand the
  robust responses of comprehensive climate models to forcing,'' {\em Climate
  Dynamics}, vol.~44, no.~7, pp.~1957--1977, 2015.

\bibitem{Rasp9684}
S.~Rasp, M.~S. Pritchard, and P.~Gentine, ``Deep learning to represent subgrid
  processes in climate models,'' {\em Proceedings of the National Academy of
  Sciences}, vol.~115, no.~39, pp.~9684--9689, 2018.

\bibitem{doi:10.1029/2018GL078202}
P.~Gentine, M.~Pritchard, S.~Rasp, G.~Reinaudi, and G.~Yacalis, ``Could machine
  learning break the convection parameterization deadlock?,'' {\em Geophysical
  Research Letters}, vol.~45, no.~11, pp.~5742--5751, 2018.

\bibitem{doi:10.1029/2018MS001390}
S.~K. Müller, E.~Manzini, M.~Giorgetta, K.~Sato, and T.~Nasuno, ``Convectively
  generated gravity waves in high resolution models of tropical dynamics,''
  {\em Journal of Advances in Modeling Earth Systems}, vol.~10, no.~10,
  pp.~2564--2588, 2018.

\bibitem{doi:10.1029/2018MS001351}
P.~A. O'Gorman and J.~G. Dwyer, ``Using machine learning to parameterize moist
  convection: Potential for modeling of climate, climate change, and extreme
  events,'' {\em Journal of Advances in Modeling Earth Systems}, vol.~10,
  no.~10, pp.~2548--2563, 2018.

\bibitem{goodfellow2014generative}
I.~Goodfellow, J.~Pouget-Abadie, M.~Mirza, B.~Xu, D.~Warde-Farley, S.~Ozair,
  A.~Courville, and Y.~Bengio, ``Generative adversarial nets,'' in {\em
  Advances in neural information processing systems}, pp.~2672--2680, 2014.

\bibitem{doi:10.1029/2019MS001896}
D.~J. Gagne~II, H.~M. Christensen, A.~C. Subramanian, and A.~H. Monahan,
  ``Machine learning for stochastic parameterization: Generative adversarial
  networks in the lorenz '96 model,'' {\em Journal of Advances in Modeling
  Earth Systems}, vol.~12, no.~3, p.~e2019MS001896, 2020.
\newblock e2019MS001896 10.1029/2019MS001896.

\bibitem{Loenze_96_GAN_2}
D.~Crommelin and W.~Edeling, ``Resampling with neural networks for stochastic
  parameterization in multiscale systems,'' 04 2020.

\bibitem{Yang2020PhysicsInformedGA}
L.~Yang, D.~Zhang, and G.~E. Karniadakis, ``Physics-informed generative
  adversarial networks for stochastic differential equations,'' {\em SIAM J.
  Scientific Computing}, vol.~42, pp.~A292--A317, 2020.

\bibitem{WU2020109209}
J.-L. Wu, K.~Kashinath, A.~Albert, D.~Chirila, Prabhat, and H.~Xiao,
  ``Enforcing statistical constraints in generative adversarial networks for
  modeling chaotic dynamical systems,'' {\em Journal of Computational Physics},
  vol.~406, p.~109209, 2020.

\bibitem{Stat_Stinis}
P.~Stinis, T.~Hagge, A.~Tartakovsky, and E.~Yeung, ``Enforcing constraints for
  interpolation and extrapolation in generative adversarial networks,'' {\em
  Journal of Computational Physics}, 03 2018.

\bibitem{Yang_Physics}
Z.~Yang and H.~Xiao, ``Enforcing deterministic constraints on generative
  adversarial networks for emulating physical systems,'' 11 2019.

\bibitem{Decoder_Based_Gen}
Y.~Wu, Y.~Burda, R.~Salakhutdinov, and R.~Grosse, ``On the quantitative
  analysis of decoder-based generative models,'' 11 2016.

\bibitem{Advar_Variational_Bayes}
L.~Mescheder, S.~Nowozin, and A.~Geiger, ``Adversarial variational bayes:
  Unifying variational autoencoders and generative adversarial networks,'' in
  {\em Proceedings of the 34th International Conference on Machine Learning -
  Volume 70}, ICML’17, p.~2391–2400, JMLR.org, 2017.

\bibitem{IntroVAE}
H.~Huang, Z.~Li, R.~He, Z.~Sun, and T.~Tan, ``Introvae: Introspective
  variational autoencoders for photographic image synthesis,'' 07 2018.

\bibitem{navier_VAE}
S.~Eismann, S.~Bartzsch, and S.~Ermon, ``Shape optimization in laminar flow
  with a label-guided variational autoencoder,'' 12 2017.

\bibitem{SupernoVAE}
X.-A. Tibau~Alberdi, C.~Requena-Mesa, C.~Reimers, J.~Denzler, V.~Eyring,
  M.~Reichstein, and J.~Runge, ``Supernovae : Vae based kernel pca for analysis
  of spatio-temporal earth data,'' 01 2018.

\bibitem{Polar_Vortex}
M.~Krinitskiy, Y.~Zyulyaeva, and S.~Gulev, ``Clustering of polar vortex states
  using convolutional autoencoders,'' 09 2019.

\bibitem{doi:10.1029/2009JD013091}
V.~P. Ghate, B.~A. Albrecht, and P.~Kollias, ``Vertical velocity structure of
  nonprecipitating continental boundary layer stratocumulus clouds,'' {\em
  Journal of Geophysical Research: Atmospheres}, vol.~115, no.~D13, 2010.

\bibitem{cifar}
A.~Krizhevsky, V.~Nair, and G.~Hinton, ``Cifar-10 (canadian institute for
  advanced research),''

\bibitem{Higgins2017betaVAELB}
I.~Higgins, L.~Matthey, A.~Pal, C.~Burgess, X.~Glorot, M.~M. Botvinick,
  S.~Mohamed, and A.~Lerchner, ``beta-vae: Learning basic visual concepts with
  a constrained variational framework,'' in {\em ICLR}, 2017.

\bibitem{Alemi2018FixingAB}
A.~A. Alemi, B.~Poole, I.~S. Fischer, J.~V. Dillon, R.~A. Saurous, and
  K.~Murphy, ``Fixing a broken elbo,'' in {\em ICML}, 2018.

\bibitem{autoencoder}
``Deep learning for universal linear embeddings of nonlinear dynamics,'' {\em
  Nature Communications}, vol.~9, no.~1, p.~4950, 2018.

\bibitem{Kingma2014AutoEncodingVB}
D.~P. Kingma and M.~Welling, ``Auto-encoding variational bayes,'' {\em CoRR},
  vol.~abs/1312.6114, 2014.

\bibitem{Bowman2016GeneratingSF}
S.~R. Bowman, L.~Vilnis, O.~Vinyals, A.~M. Dai, R.~J{\'o}zefowicz, and
  S.~Bengio, ``Generating sentences from a continuous space,'' in {\em CoNLL},
  2016.

\bibitem{10.1175/1520-0469(2003)060<0607:CRMOTA>2.0.CO;2}
M.~Khairoutdinov and D.~Randall, ``Cloud resolving modeling of the arm summer
  1997 iop: Model formulation, results, uncertainties, and sensitivities,''
  {\em Journal of The Atmospheric Sciences - J ATMOS SCI}, vol.~60,
  pp.~607--625, 02 2003.

\bibitem{Khairoutdinov1999ALE}
M.~Khairoutdinov and Y.~L. Kogan, ``A large eddy simulation model with explicit
  microphysics: Validation against aircraft observations of a
  stratocumulus-topped boundary layer,'' 1999.

\bibitem{10.1175/MWR3467.1}
A.~Clark, W.~Gallus, and T.-C. Chen, ``Comparison of the diurnal precipitation
  cycle in convection-resolving and non-convection-resolving mesoscale
  models,'' {\em Monthly Weather Review - MON WEATHER REV}, vol.~135, 10 2007.

\bibitem{10.5555/2380985}
K.~P. Murphy, {\em Machine Learning: A Probabilistic Perspective}.
\newblock The MIT Press, 2012.

\bibitem{Cooley1965AnAF}
J.~W. Cooley and J.~W. Tukey, ``An algorithm for the machine calculation of
  complex fourier series,'' 1965.

\bibitem{pearson_karl_1901_1430636}
K.~Pearson, ``{LIII. On lines and planes of closest fit to systems of points in
  space},'' Nov. 1901.

\end{thebibliography}

\end{document}